\documentclass[aps,reprint,longbibliography]{revtex4-2}
\usepackage{amsmath}
\usepackage{enumerate}
\usepackage[utf8]{inputenc} 
\usepackage[T1]{fontenc} 
\usepackage{lmodern} 
\usepackage[colorlinks=true, linkcolor=blue, citecolor=blue, urlcolor=blue]{hyperref}
\usepackage{bbm}
\usepackage{graphicx}
\usepackage{color}
\usepackage{orcidlink}
\usepackage{mathtools}
\usepackage{bm}
\usepackage{txfonts}
\usepackage{soul}
\def\bu{\mathbf{u}}
\def\bv{\mathbf{v}}
\def\br{\mathbf{r}}
\def\bk{\mathbf{k}}
\def\uu{\mathbf{u}}
\def\vv{\mathbf{v}}
\def\x{\mathbf{x}}
\newcommand{\ip}[2]{\langle#1|#2\rangle}
\DeclareMathOperator{\sinc}{sinc}

\def\dwc{Dodd-Walls Centre for Photonic and Quantum Technologies, Dunedin 9054, New Zealand}
\def\uop{Department of Physics, University of Otago, Dunedin 9016, New Zealand}
\newcommand{\eref}[1]{(\ref{#1})}
\newcommand{\eeref}[1]{Eq.~(\ref{#1})}
\newcommand{\fref}[1]{Fig.~\ref{#1}}
\begin{document}
\title{Velocity correlations of vortices and rarefaction pulses in compressible planar quantum fluids}
\author{Ashton S. Bradley \orcidlink{0000-0002-3027-3195}}
\email{ashton.bradley@otago.ac.nz}
\affiliation{\dwc}
\affiliation{\uop}
\author{Nils A. Krause \orcidlink{0009-0009-0754-7782}}
\affiliation{\dwc}
\affiliation{\uop}
\date{\today}

\begin{abstract}
We present a quantitative analytical framework for calculating two-point velocity correlations in compressible quantum fluids, focusing on two key classes of superfluid excitations: vortices and rarefaction pulses. We employ two complementary approaches. First, we introduce a new ansatz for vortex cores in planar quantum fluids that enhances analytic integrability. This ansatz yields closed-form expressions for power spectra and velocity correlations for general vortex distributions. Using it, we identify distinct signatures of short- and long-range velocity correlations corresponding to vortex dipoles and vortex pairs, respectively. Second, we analyze the fast rarefaction pulse regime of the Jones-Roberts soliton. By applying the asymptotic high-velocity wavefunction, we derive analytic expressions for the velocity power spectrum and correlation function, capturing the soliton characteristic length scale. We validate our analytical results for the homogeneous system against numerical treatment of a large trapped system, finding close quantitative agreement. Our findings provide the first analytic treatment of velocity correlations for Jones-Roberts solitons in quantum fluids of light [M. Baker-Rasooli et al., Physical Review Letters, 134, 233401 (2025)], and establish a foundation for characterizing vortices and solitons in compressible quantum fluids.
\end{abstract}
\maketitle
\section{Introduction}
Dilute-gas Bose-Einstein condensates (BECs), as canonical example quantum fluids, are increasingly being explored in far from equilibrium regimes. Recent experiments in planar BEC systems~\cite{gauthier_giant_2019,johnstone_evolution_2019,gazo_universal_2023,karailiev_observation_2024, geng_rayleigh-taylor_2024} have significantly deepened our understanding of compressible quantum fluids, and their connections to classical incompressible counterparts. Compressible quantum fluids---including atomic BECs, polariton condensates~\cite{deng_condensation_2002}, and quantum fluids of light~\cite{fontaine_observation_2018} (QFL)---can support both incompressible quantum vortices~\cite{pitaevskii_vortex_1961,matthews_vortices_1999,madison_vortex_2000,lagoudakis_quantized_2008} with finite core scale, and compressible excitations such as sound waves. These excitations can interconvert via vortex dipole decay or spontaneous vortex formation. A particlar  compressible excitation forms in the adiabatic decay of a vortex dipole, a two-dimensional quasi-soliton of Jones-Roberts type~\cite{jones_motions_1982,meyer_observation_2017}---known as a rarefaction pulse---that carries both energy and momentum.

In planar systems vortices can store significant energy by clustering, exhibiting negative temperature~\cite{onsager_statistical_1949,gauthier_giant_2019,johnstone_evolution_2019}, and high energy velocity fields. Beyond the s-wave interacting scalar BEC, more complex systems---such as spinor BEC~\cite{kawaguchi_spinor_2012}, spin-orbit coupling~\cite{lin_spin-orbit-coupled_2011}, and dipolar interactions~\cite{bland_vortices_2023}---exhibit a range of exotic behaviors including anisotropic superfluidity~\cite{ticknor_anisotropic_2011,mulkerin_anisotropic_2013}, and supersolidity~\cite{casotti_observation_2024}, associated with non-trivial superfluid dynamics. Yet characterisation of superfluid motion in BEC has significant limitations: simultaneous access to both density and phase of the superfluid order parameter is challenging, critical information for analysis of quantum turbulence~\cite{nore_kolmogorov_1997,bradley_spectral_2022}. However, in quantum fluids of light this information is directly accessible~\cite{glorieux_hot_2023}, enabling the study of superfluid dynamics with high spatio-temporal resolution~\cite{baker-rasooli_turbulent_2023}.

In this work we develop a treatment of energy spectra and velocity correlations applicable to vortices and rarefaction pulses in compressible quantum fluids. We introduce a new exponential ansatz for the vortex core of a Gross-Pitaevskii fluid, with advantages for analysis in the Fourier domain. We use it to construct velocity power spectra and two-point correlation functions for general vortex distributions in the homogeneous fluid, including a finite size cutoff. The ansatz has improved analytical integrability properties in 2D, yielding analytical results for velocity correlations first presented in \cite{baker-rasooli_observation_2025}. 

To gain insight as to the link between vortex distributions and velocity correlations length scales, we consider few-vortex states including the single vortex, vortex pair, and vortex dipole. We use an incompressible approximation that simplifies the resulting correlation functions. We also consider the rarefaction pulse regime of the Jones-Roberts soliton~\cite{jones_motions_1982,krause_thermal_2024}, the high velocity counterpart of a vortex dipole. We derive an analytical expression for the velocity correlation function of a rarefaction pulse in a homogeneous superfluid, in the high velocity regime~\cite{tsuchiya_solitons_2008,krause_thermal_2024}. We compare our analytical results to numerical results for excitations introduced in the center of a BEC confined by a harmonic trap. The velocity correlation function gives a clear signature of length scales associated with these compressible and incompressible excitations, providing a useful tool for characterizing quantum fluids. 

As vortex dipoles and rarefaction pulses are part of a family of nonlinear GPE eigenstates carrying linear momentum, our work provides a treatment of velocity correlations for these low-energy excitations in compressible quantum fluids. Our approach is also motivated by recent experimental work on quantum fluids of light~\cite{baker-rasooli_observation_2025}, and is relevant to the study of quantum turbulence in atomic BECs~\cite{neely_characteristics_2013,navon_emergence_2016,gauthier_giant_2019,johnstone_evolution_2019,fischer_regimes_2025}. 

This paper is structured as follows. In Sec.~\ref{sec:background} we provide background on the Gross-Pitaevskii equation and the properties of vortices and rarefaction pulses. In Sec.~\ref{sec:vortices} we outline our analytical methods for calculating velocity correlations for distributions of quantum vortices. In Sec.~\ref{sec:rarefaction} we present our analytical treatment of the rarefaction pulse regime. In Sec.~\ref{sec:trap} we compare the homogeneous theory to numerical results for incompressible and compressible excitations of a BEC in a harmonic trap. Finally, in Sec.~\ref{sec:conclusions} we conclude and discuss future directions.

\section{Background}
\label{sec:background}
Several experimental measures are used to characterize atomic BECs, and a range of techniques are employed to analyse experiment and simulation data. Experimental approaches include time of flight absorption imaging to approximate the momentum distribution~\cite{anderson_observation_1995}, interference~\cite{andrews_observation_1997}, and Bragg spectroscopy~\cite{stenger_bragg_1999-1}. However the simultaneous measurement of the density and phase of the BEC order parameter is ruled out by atom number conservation. Furthermore, while Bragg scattering can yield vortex locations and charges~\cite{seo_observation_2017}, technical challenges have limited widespread adoption of the technique. The recent development of particle image velocimetry~\cite{zhao_kolmogorov_2025} offers a promising approach to addressing these challenge in BEC experiments. In theoretical studies of compressible quantum turbulence, kinetic energy analysis using Helmholtz decomposition of a weighted velocity field~\cite{nore_kolmogorov_1997} allows singularity-free separation of compressible, incompressible, and quantum pressure components of the fluid kinetic energy~\cite{bradley_energy_2012,bradley_spectral_2022}. However, due to limited density and velocity information this analysis has not been applied to BEC experiments. In contrast, in QFL density and phase information is directly accessible~\cite{glorieux_hot_2023}, enabling high-precision characterisations of superfluid dynamics~\cite{baker-rasooli_turbulent_2023}.

Kinetic energy spectra, and their corresponding two-point correlation functions, provide useful measures of superfluid properties~\cite{nore_kolmogorov_1997,naraschewski_spatial_1999,bradley_spectral_2022}. These quantities are useful for characterizing the length scales of excitations in the fluid, providing signatures of vortices, rarefaction pulses, and other excitations, and for understanding the physical origins of power-law behavior in quantum turbulence~\cite{navon_emergence_2016,fischer_regimes_2025}. While spectral representations can be useful for identifying power laws in scale-invariant states, well-defined length scales in the system are often more easily identifiable as correlation length scales in the spatial domain. The velocity correlations associated with excitations such as distributions of vortices and rarefaction pulses can shed light on general properties of quantum fluids, and quantum turbulent states~\cite{neely_characteristics_2013,navon_emergence_2016,gauthier_giant_2019,johnstone_evolution_2019,fischer_regimes_2025}. 

As a canonical excitation of planar quantum fluids, the Jones-Roberts soliton (JRS), consists of a vortex-antivortex pair (vortex dipole) at low velocity (high energy), through to a localised density dip (rarefaction pulse) at high velocity (low energy)~\cite{jones_motions_1982,jones_motions_1986}. JRS are strictly quasi-solitons, forming the planar superfluid analogue of the one-dimensional dark soliton~\cite{burger_dark_1999}. As the energy in a vortex dipole is entirely incompressible, while a rarefaction pulse is entirely compressible, the JRS provide an interesting system in which to study velocity power spectra and spatial correlation functions, motivating the analytical treatment presented in this work.

Theoretical treatment of compressible vortices is complicated by the vortex core structure~\cite{fetter_vortices_2001,bradley_energy_2012,mehdi_mutual_2023}. Despite some progress in spectral representation~\cite{bradley_energy_2012}, the commonly used vortex core ansatz~\cite{fetter_vortices_2001} limits analytical results. In particular, an inversion of vortex spectra back to position space to find correlation functions~\cite{bradley_spectral_2022} is not known analytically. In a recent experimental and theoretical work the velocity correlation function of compressible and incompressible excitations in a quantum fluid of light were studied in detail~\cite{baker-rasooli_observation_2025}, enabled by high resolution measurements and spectral analysis~\cite{bradley_spectral_2022}, and a new analytical approach to the vortex core that we present in detail in this article. 

In the remainder of this section we briefly review the Gross-Pitaevskii equation (GPE) and the properties of quantum vortices and rarefaction pulses, and introduce the velocity power spectrum. 
\subsection{Gross-Pitaevskii equation}
We consider a system described by the Gross-Pitaevskii equation (GPE) in 2D
\begin{align}
    i\hbar\frac{\partial\psi(\br,t)}{\partial t}&=\left(-\frac{\hbar^2}{2m}\nabla^2+V(\br)+\varg|\psi(\br,t)|^2\right)\psi(\br,t),
\end{align}
where $\psi(\br,t)$ is the complex wavefunction, $V(\br)$ is the external potential, and $\varg$ is the two-body  interaction strength. We consider the homogeneous system, $V(\br)=0$, and the fluid is in the superfluid phase, with a condensate wavefunction $\psi(\br,t)=\sqrt{n(\br,t)}e^{i\Theta(\br,t)}$. The density $n(\br,t)=|\psi(\br,t)|^2$ and the velocity field $\bv(\br,t)=\hbar/m\nabla\Theta(\br,t)$. This system supports quantum vortices, topological defects in the phase of the wavefunction~\cite{fetter_vortex_2001} that carry angular momentum. It also supports rarefaction pulses, which are compressible solitary waves that move at a constant velocity~\cite{jones_motions_1982,jones_motions_1986} and carry linear momentum. 

A homogeneous system with background density $n_0$ has natural length unit given by the healing length $\xi$, related to chemical potential $\mu$ and speed of sound $c$ via
\begin{align}
    \mu = gn_0 = \frac{\hbar^2}{m\xi^2}=mc^2.
\end{align} 
\subsection{Quantum vortices}
Vorticity is quantized for the single-valued wavefunction, and the circulation around any closed path $\mathcal{C}$ is given by
\begin{align}
    \oint_{\mathcal{C}}\bv\cdot d\br = \frac{\hbar}{m}\oint_{\mathcal{C}}\nabla\Theta\cdot d\br = \frac{h}{m} q,
\end{align}
where $q=\pm 1$ is the circulation quantum. 
The vortex velocity field is
\begin{align}
    \bv(\br) &= \frac{\hbar}{mr} \begin{pmatrix}
        -\sin\theta\\
        \cos\theta
    \end{pmatrix},
\end{align}
and this field is curl free except precisely at the vortex core, where $\nabla\times\bv(\br) = 2\pi q\delta(\br)$. The vortex core is a region of suppressed density where the phase winds by $2\pi$, and the core size is of order the healing length $\xi$. In general the exact vortex core shape must be found numerically~\cite{fetter_vortices_2001}.

\subsection{Vortex core shape ansatz}
Vortex core shape approximations have a long history in the study of quantum vortices. The simplest is the point vortex approximation, where the core is taken to be a delta function. An ansatz for the vortex core profile was introduced by Fetter~\cite{fetter_vortices_1965} and was since applied to trapped BEC~\cite{fetter_vortices_2001}, and a range of vortex dynamics problems~\cite{pethick_bose-einstein_2008,fialko_quantum_2012}.

A particular choice of the Fetter ansatz that matches the slope at the origin was also introduced to calculate analytical expressions for velocity power spectra~\cite{bradley_energy_2012}. The vortex wavefunction was assumed to take the form
\begin{align}\label{chia}
    \psi_F(\br) &\equiv \sqrt{n_0}\frac{r}{\sqrt{r^2+(\xi/\Lambda)^2}} e^{\pm i\theta},
\end{align}
where
\begin{align}
    \Lambda = \lim_{r\to 0} \frac{d}{d(r/\xi)}\log{\psi_v(\br)}= 0.8249\dots,
\end{align}
is the slope of the exact vortex wavefunction at the vortex core. This ansatz reproduces the correct slope at the origin, the asymptotic behaviour at small and large radii, and the vortex velocity field, and is thus well-suited for spectral analysis.
\subsection{Rarefaction pulse}
The Jones-Roberts soliton is a compressible solitary wave solution of the GPE that moves at a constant velocity $v$~\cite{jones_motions_1982}. For $0.8c\lesssim v\leq c$ it has the asymptotic form~\footnote{There is a lower bound due to the requirement that the density remain positive.}
\begin{align}
    n(\br) &= n_0 \left(1 - 4\epsilon^2 \frac{3/2 + 2(\epsilon^2 y/\xi)^2 - 2(\epsilon x/\xi)^2}{[3/2 + 2(\epsilon^2 y/\xi)^2 + 2(\epsilon x/\xi)^2]^2}\right), \label{jrn}\\
    \Theta(\br) &= -2\sqrt{2}\epsilon \frac{\sqrt{2}\epsilon x/\xi}{3/2 + 2(\epsilon^2 y/\xi)^2 + 2(\epsilon x/\xi)^2},\label{jrp}
\end{align}
where $\epsilon = \sqrt{1 - (v/c)^2}$ and $c$ is the speed of sound in the fluid. The density $n(\br)$ and phase $\Theta(\br)$ describe the rarefaction pulse regime of a Jones-Roberts soliton, with factors of $\sqrt{2}$ due to our definition of $\xi$.

\subsection{Velocity power spectrum}
The velocity power spectrum $E(k)$ of a compressible quantum fluid in two dimensions is defined as the momentum space kernel of the kinetic energy integral 
\begin{align}\label{Eall}
    E&=\frac{m}{2}\int d^2\br\; n(\br)|\vv|^2 =\int_0^\infty E(k)dk.
\end{align}
Parseval's theorem may be used to show
\begin{align}
    E(k)&= \frac{m}{2}k\int_0^{2\pi}d\phi\: |\tilde\bu(\bk)|^2,
    \label{vspecdef}
\end{align}
where $\phi$ is the polar angle in $k$-space, and the Fourier transform is
\begin{align}
    {\cal F}[f](\bk)&=\frac{1}{2\pi}\int d^2\br\:e^{-i\bk\cdot\br}f(\br).
\end{align} 
The density weighted velocity field $\bu(\br)\equiv\sqrt{n(\br)}\bv(\br)$ is used to define the power spectrum for a compressible fluid as it is required to regularize the velocity singularity at a vortex core. 

The vector field $\bu(\br)$ may be further decomposed into its compressible (curl free) and incompressible (divergence free) components using a Helmholtz decomposition, $\bu(\br)=\bu^c(\br)+\bu^i(\br)$, where $\nabla\cdot\bu^i=0$ and $\nabla\times\bu^c=0$~\cite{nore_kolmogorov_1997}. The decomposition may be efficiently performed in momentum space~\cite{bradley_spectral_2022}, and power spectra may be constructed using an angle-integrated Wiener-Khinchin theorem~\cite{bradley_spectral_2022} as described in Appendix \ref{app:c}. Either by the replacement $\uu\to \uu^\alpha$ in \eref{Eall}, \eref{vspecdef}, or by taking $\uu=\vv=\uu^\alpha$ in the appendix, we arrive at the power spectral densities and total energies related by
\begin{align}
    E^\alpha&\equiv\int_0^\infty dk\;E^\alpha(k)
\end{align}
for incompressible and compressible components of the kinetic energy denoted by $\alpha\in\{i,c\}$ respectively.

In the present work it is important to emphasize that a vortex-free field has no incompressible part in the above decomposition, allowing analytical attention on the complete velocity field for the case of the rarefaction pulse. Similarly, for the ideal well-separated vortex systems of interest in this work, it is a very good approximation to ignore compressible effects for the purpose of analytical work hence do not need to confront the decomposition as part of the analysis. Further discussion of this approximation is given in Ref.~\cite{bradley_energy_2012}.

\subsection{Vortex power spectrum}
It is useful to give a short summary of the incompressible velocity power spectrum for a single vortex, independent of the particular core shape used for specific calculations. We keep the derivation as general as possible with respect to the vortex core shape.

In general we can define 
\begin{align}
    \psi(\br)&\equiv\sqrt{n_0}\chi(r/\xi)e^{\pm i\theta}.
\end{align}
For any core shape $\chi(r/\xi)$, assuming a pure azimuthal velocity field, to determine the power spectrum, we want to find the two dimensional Fourier transform of 
\begin{align}
    \bu_s(\br)&=\sqrt{n(\br)}\bv(\br)=\frac{\hbar\sqrt{n_0}}{m}\frac{\chi(r/\xi)}{r}\begin{pmatrix}
        -\sin\theta\\
        \cos\theta
    \end{pmatrix}.
    \label{svort}
\end{align}
In the Fourier domain
\begin{align}
    \tilde\bu_s(\bk)&\equiv\frac{1}{2\pi}\int d^2\br\:e^{-i\bk\cdot\br}\bu_s(\br)\notag\\
    &=\frac{\hbar\sqrt{n_0}}{m}\int_0^\infty dr\:\chi(r/\xi)\frac{1}{2\pi}\int_0^{2\pi}d\theta\:e^{ikr\cos{(\theta-\phi)}}\begin{pmatrix}
        -\sin\theta\\
        \cos\theta
    \end{pmatrix},
\end{align}
where $\phi$ is the angle $\bk$ makes to the $x$ axis. Using the Bessel function identity
\begin{align}
    \frac{1}{2\pi}\int_0^{2\pi}d\theta\:e^{ikr\cos{(\theta-\phi)}}\begin{pmatrix}
        -\sin\theta\\
        \cos\theta
    \end{pmatrix} &=iJ_1(kr)\begin{pmatrix}
        -\sin\phi\\
        \cos\phi
    \end{pmatrix},
\end{align}
and, provided $k>0$ we have 
\begin{align}
    J_1(kr)=-\frac{1}{k}\frac{dJ_0(kr)}{dr}.
\end{align} 
We can integrate by parts and change variables to find
\begin{align}\label{utilde}
    \tilde\bu_s(\bk)&=\frac{i\hbar\sqrt{n_0}}{mk}\int_0^\infty d\sigma\:\chi'(\sigma)J_0(k\xi \sigma)\begin{pmatrix}
        -\sin\phi\\
        \cos\phi
    \end{pmatrix},
\end{align}
a convenient form, since $\chi'(\sigma)\to 0$ as $\sigma\to\infty$. The incompressible velocity power spectrum is now found from \eref{vspecdef}, \eref{utilde} as
\begin{align}
    E(k) & = \varepsilon_0\xi F(k\xi),\label{onevE}
\end{align}
where 
\begin{align}
    F(z)&\equiv\frac{1}{z}\left(\int_0^\infty d\sigma\:\chi'(\sigma)J_0(\sigma z)\right)^2,
    \label{Fdef}
\end{align}
and 
\begin{align}\label{e0def}
    \varepsilon_0&\equiv\frac{\pi\hbar^2n_0}{m}=\pi \mu n_0\xi^2
\end{align} 
is the energy unit.
In this form we can consider different ansatz for single or multiply charged cores, dipolar vortices, etc. 

The function $F(z)$ has the following asymptotic properties:
\begin{enumerate}[(i)]
    \item Point vortex limit: $F(z)\to 1/z$ for $z\ll 1$, independent of the particular core shape. This follows since $\chi'$ is only nonzero over a small range of $\sigma\lesssim 10$, and so for small $z$, $J_0(\sigma z)\simeq J_0(0)=1$ and may be taken outside the integral. Since $\chi(0)=0$ and $\chi(\infty)=1$ we have
    \begin{align}
        \int_0^\infty \chi'(\sigma)d\sigma&=-1.
    \end{align}
    \item Vortex core limit: $F(z)\to \Lambda^2/z^3$ for $z\gg 1$. This follows from the change of variables $u = \sigma z$, giving
    \begin{align}
        \int_0^\infty d\sigma\:\chi'(\sigma)J_0(\sigma z)&=\frac{1}{z}\int_0^\infty du\:\chi'(u/z)J_0(u)\to\frac{\chi'(0)}{z},
    \end{align}
\end{enumerate}
since $\int_0^\infty du\:J_0(u)=1$. 

We have summarized key background material required to develop the theory of velocity correlations for vortices and rarefaction pulses in compressible quantum fluids.  Note that for any particular vortex core function $\chi(r/\xi)$, the velocity power spectrum can be evaluated via \eref{onevE},\eref{Fdef}. 
\section{Vortices}
\label{sec:vortices}
To proceed further analytically, we explore the properties of a new vortex core ansatz function.
In previous work, a particular form of the spectrum was calculated using the Fetter ansatz in \cite{bradley_energy_2012}. It had the correct UV and IR asymptotics, but the mid range involved a product of Bessel functions, limiting further analytical work. In this section we calculate the velocity power spectrum and two-point velocity correlation function for a system of compressible quantum vortices in a homogeneous fluid. For this purpose we introduce a new ansatz for the vortex core that simplifies the calculation of the power spectrum and velocity correlations for general vortex distributions.

\subsection{Exponential ansatz}
For the purpose of analytical work, we seek a description of the compressible fluid vortex core that is more tractable than the Fetter ansatz, \eeref{chia}. The key property we seek is a simpler Fourier transform, with the final aim that a Fourier inversion of the velocity power spectrum can be performed analytically to find the two-point velocity correlation function.

\begin{figure}[!t]
    \centering
    \includegraphics[width=\columnwidth]{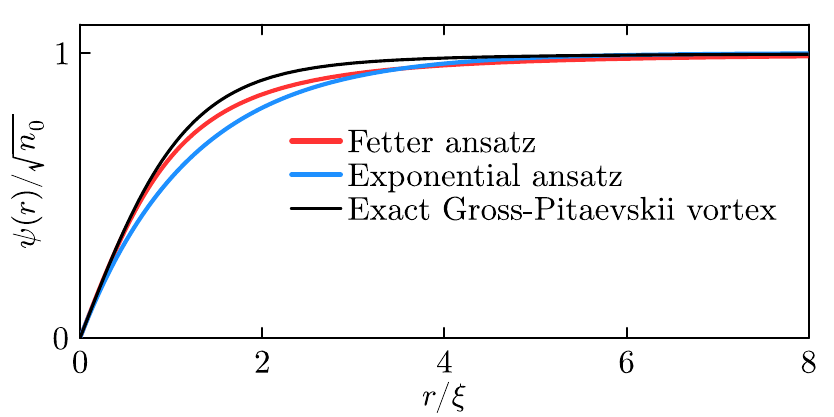}
    \caption{Comparison of the Fetter [\eeref{chia}] and exponential [\eeref{eansatz}] ans\"{a}tze with the exact core shape. The black line shows the numerically exact core solution for the two-dimensional Gross-Pitaevskii equation. The blue line shows the exponential ansatz~\eref{chiedef}, and the red line shows the Fetter ansatz \eref{chia}.}
    \label{fig:ansatz}
\end{figure}

We introduce the following exponential ansatz
\begin{align}\label{eansatz}
    \psi_\Lambda(\mathbf{r}) &\equiv \sqrt{n_0}(1-e^{-\Lambda r/\xi})e^{\pm i\theta},
\end{align}
with the correct vortex slope at the origin, and the correct asymptotic for a GPE vortex at large radii. The exponential ansatz \eref{eansatz} is compared with the Fetter ansatz and the exact core shape in Fig.~\ref{fig:ansatz}. We note that the exponential ansatz converges to the exact far field density more rapidly than the Fetter ansatz. However, the core shape is not as accurate, but approaches the correct slope $\Lambda/\xi$ for $r\ll\xi$. Note also that both $\psi_
\Lambda$ and $\psi_F$ satisfy $\nabla\cdot(n\vv)=0$, the time independent continuity equation, since the velocity field is azimuthal and rotationally invariant.

\subsection{Vortex power spectrum}
The form \eref{eansatz} has core shape
\begin{align}
    \chi_\Lambda(\sigma)&\equiv(1-e^{-\Lambda \sigma}).
    \label{chiedef}
\end{align}
To evaluate \eeref{Fdef} for this core, we use \eref{appj1} to find
\begin{align}
    \int_0^\infty d\sigma\:\chi_\Lambda'(\sigma)J_0(\sigma z)&=\frac{\Lambda}{\sqrt{z^2+\Lambda^2}},
\end{align}
and the spectral function
\begin{align}
    F_\Lambda(z)&=\frac{\Lambda^2}{z(z^2+\Lambda^2)}=\frac{1}{z}-\frac{z}{z^2+\Lambda^2}.
    \label{Fk}
\end{align}
This representation of the vortex core transform has properties $F_\Lambda(z)\to 1/z$ as $z\to 0$, and $F_\Lambda(z)\to \Lambda^2/z^3$ as $z\to \infty$, giving spectrum
\begin{align}
    E^i(k)&=\varepsilon_0\xi F_\Lambda(k\xi),
    \label{Ekvort}
\end{align}
with a point-vortex term dominant at wavenumbers $2\pi/R \ll k\ll 2\pi/\xi$, and a core contribution evident for large wave numbers $k\gg 2\pi/\xi$, as shown in Fig.~\ref{fig:correlator} (a). 

The spectrum captures the behavior at small and large scales. However, the $1/z$ behavior at small $z$ leads to finite-size divergences in the velocity field, and hence in the two-point velocity correlation function. To regularize this we introduce an infrared cutoff $\Gamma=2\pi/\bar{R}$, where $\bar{R}=R/\xi\gg 1$ is the system size. This can be imposed in the spectral function in a tractable way by introducing the regularized core   
\begin{align}
    F_\Gamma(z)&\equiv\frac{1}{\sqrt{z^2+\Gamma^2}}-\frac{z}{z^2+\Lambda^2},
    \label{Fgam}  
\end{align}
so that $F_\Gamma(z)\to F_\Lambda(z)$ as $R\to\infty$. The single vortex spectrum is then 
\begin{align}\label{Egam_one}
    E^i_\Gamma(k)&=\varepsilon_0\xi F_\Gamma(k\xi).
\end{align}
The net effect is to remove the infrared divergence in the velocity field of the vortex, as shown in Fig.~\ref{fig:correlator} (a).
\subsection{Two-point velocity correlation function}
A useful measure of the velocity field is the two-point velocity correlation function, which describes the correlation between the velocity field at two points separated by a distance $r$. The point-wise correlation is averaged over the system to give a global statistical measure of the velocity field correlation length scales. The correlation function is also calculated as the Fourier transform of the velocity power spectrum. 

In detail, the two-point velocity correlation function between two vector fields is defined as
\begin{align}
    C[\bu,\bv](\br) &\equiv \int d^2\mathbf{R}\:\bu^*(\mathbf{R}-\br/2)\cdot \bv(\mathbf{R}+\br/2)
\end{align}
and we are interested in the angle-averaged autocorrelation function 
\begin{align}
    G(r) &\equiv \frac{1}{2\pi}\int_0^{2\pi} d\theta\:C[\bu,\bu](r\cos\theta,r\sin\theta).
\end{align}
This two-point velocity correlation may also be calculated via the inverse Fourier transform of the velocity power spectrum~\footnote{This is nothing more than an application of the Wiener-Khinchin theorem under angular integration.}, as the integral~\cite{bradley_spectral_2022}
\begin{align}
    G(r)&=\int_0^\infty dk\:E(k)J_0(kr).
    \label{Gdef}
\end{align}
The simplicity of the expression \eref{Ekvort} allows us to evaluate $G(r)$ within the exponential ansatz, using \eref{Fk}, \eref{Ekvort}, 
\begin{align}
    G_\Lambda^i(r)&=\varepsilon_0\xi\int_0^\infty dk\:F_\Lambda(k\xi)J_0(kr)\\\notag
    &=\varepsilon_0\int_0^\infty dz\:J_0(zr/\xi)\left(\frac{1}{z}-\frac{z}{z^2+\Lambda^2}\right).
\end{align}
The second term can be easily evaluated using \eref{appj2}, however, in the first term there is an apparent divergence at $z=0$ due to the long range velocity field. Fortunately, the regularized form \eref{Fgam} is integrable:
\begin{align}\label{regint}
    \int_0^\infty dz\:\frac{1}{\sqrt{z^2+\Gamma^2}}J_0(zx)&=I_0\left(\frac{\Gamma x}{2}\right) K_0\left(\frac{\Gamma x}{2}\right),
\end{align}
where $I_0(x)$ and $K_0(x)$ are the modified Bessel functions of the first and second kind respectively. 

\begin{figure}[!t]
    \centering
    \includegraphics[width=\columnwidth]{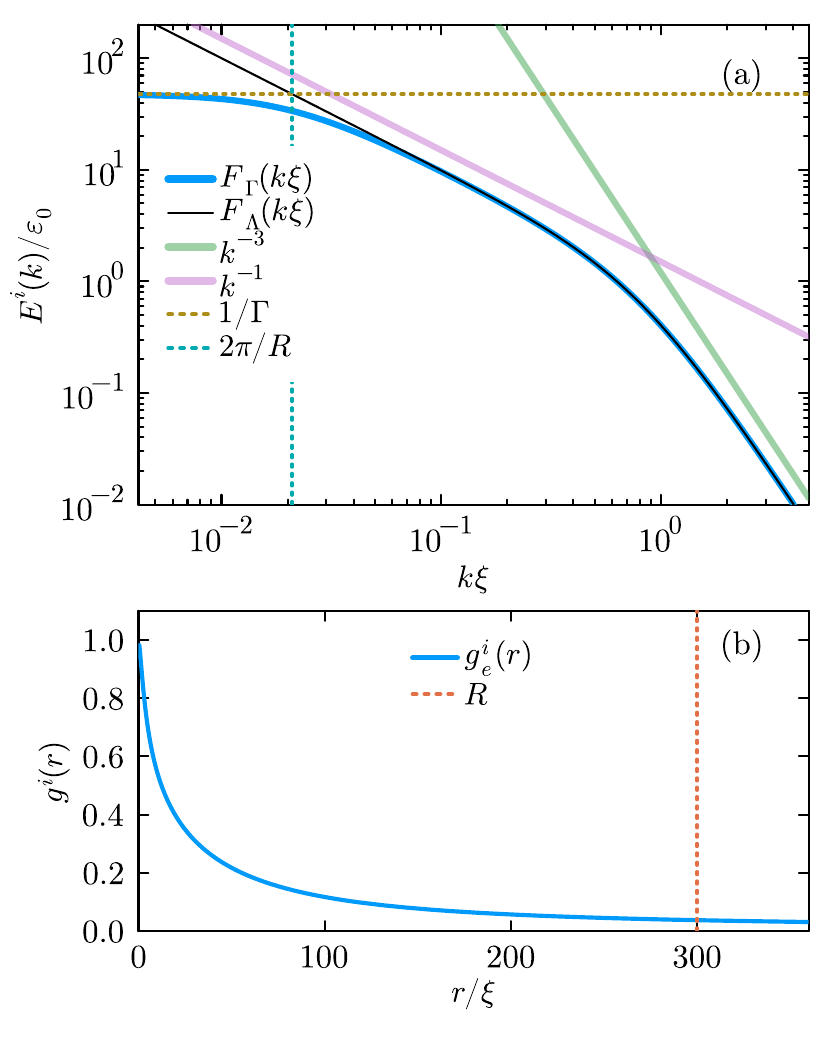}
    \caption{Single vortex properties. (a) Incompressible velocity power spectrum. Blue curve shows the IR regularized exponential for $R=300\xi$. Spectrum shows the expected vortex power laws for the near and far field, until $k\xi\sim \Gamma=2\pi/R$ where it flattens due to the infrared regularisation. (b) Normalized two-point velocity correlation for the exponential ansatz, corresponding to the spectrum in (a).}
    \label{fig:correlator}
\end{figure}

Carrying out the inversion integral~\eref{Gdef}, for \eref{Egam_one} we find the two-point correlator 
\begin{align}
    G_\Gamma^i(r)&=\varepsilon_0\left(K_0\left(\frac{\Gamma r}{2\xi}\right)I_0\left(\frac{\Gamma r}{2\xi}\right)-K_0\left(\frac{\Lambda r}{\xi}\right)\right).
    \label{Ge}
\end{align}
Since $K_0(x)I_0(x)-K_0(ax)\to \ln(a)$ as $x\to 0$, we have
the normalized incompressible velocity correlation function for a single vortex 
\begin{align}
    g_\Gamma^i(r)&\equiv\frac{G^i_\Gamma(r)}{G^i_\Gamma(0)}=\frac{K_0\left(\frac{\Gamma r}{2\xi}\right)I_0\left(\frac{\Gamma r}{2\xi}\right)-K_0\left(\frac{\Lambda r}{\xi}\right)}{\ln\left(\frac{2\Lambda}{\Gamma}\right)},
    \label{nvcorr}
\end{align}
shown in Fig.~\ref{fig:correlator} (b).
The singular behavior at the origin in the first term is removed by the second term, reliant on the separation of scales $\Gamma\ll\Lambda$, equivalent to $\xi\ll R$, always valid in the point-vortex regime where the background superfluid density is large. For $z\gg 1$, the second term is exponentially suppressed and $I_0(z)K_0(z)\to 1/(2z)$ at leading order, and hence the correlator decay is akin to that of the single-vortex velocity field. 

The expression \eref{nvcorr} is our main result. It provides an accurate analytical treatment of velocity two-point correlations for a single vortex in a finite homogeneous fluid, and is valid for all $r\geq 0$. As we show below, it can be used to develop the treatment of distributions of vortices, with high accuracy provided they are well-separated compared to the healing length. 

\subsection{Vortex distributions}
We now extend this approach to a system of $N$ well-separated vortex cores with sign of circulation $q_i\in\{+1,-1\}$, and position $\br_i$. For separated cores the phase becomes additive and we can convolve the point-vortex vorticity 
\begin{align}\label{pvdist}
    \omega(\br)&=\sum_{i}q_i\delta(\br-\br_i)
\end{align} 
with the single vortex velocity field
\begin{align}
    \bu(\br)&=\int d^2\br'\bu_s(\br-\br')\omega(\br')\\
    &=\sum_i q_i\bu_s(\br-\br_i).\label{udefs}
\end{align}
The incompressible velocity power spectrum is then found from~\eref{vspecdef} using \eref{udefs}, \eref{Fgam} 
\begin{align}
    E^i(k)&=\frac{mk}{2}\int_0^{2\pi}d\phi\:\sum_{i,j}q_iq_j e^{i\bk\cdot(\br_i-\br_j)}|\tilde\bu_s(\bk)|^2
\end{align}
to find
\begin{align}
    E^i(k)&=\varepsilon_0\xi F_\Gamma(k\xi)\sum_{i=1,j=1}^Nq_iq_jJ_0(k|\br_i-\br_j|),
    \label{Eivort}
\end{align}
akin to the result found in ~\cite{bradley_energy_2012} but with $F_\Gamma(z)$ now including both the vortex core and the infrared regularization in a form amenable to further analytical treatment. It is important to emphasize that when vortices are closer than a few healing lengths, this treatment will no longer be complete and significant modifications to the power spectrum can be expected. The core-separable form \eref{Eivort} cannot describe mutual distortion occurring for very close vortex approaches in high-energy vortex clustering, or dipole annihilation~\cite{bradley_spectral_2022,reeves_inverse_2013}. 

To calculate the two-point velocity correlator~\eref{Gdef} we need to evaluate the generalization of \eeref{Gdef} in the form 
\begin{align}
    G(a,b)&\equiv\int_0^\infty dz\:F_\Gamma(z)J_0(za)J_0(zb).
\end{align}
We evaluate this integral in Appendix \ref{appU}, arriving at the result
\begin{align}
G(a,b)&=I_0\left(\frac{\Gamma \alpha}{2}\right)K_0\left(\frac{\Gamma \alpha}{2}\right)-I_0(\Lambda \beta )K_0(\Lambda \alpha ),\label{Uex}
\end{align}
where $\alpha\equiv\max(a,b)$ and $\beta\equiv\min(a,b)$, and we note that $G(a,b)=G(b,a)$. This enables a singularity-free construction of the two-point velocity correlator for a general multivortex system. 

The general two-point correlator can now be written as
\begin{align}
    G^i(r)&=\varepsilon_0\xi\sum_{i=1,j=1}^Nq_iq_jG(r/\xi, r_{ij}/\xi), 
\end{align}
where $r_{ij}=|\br_i-\br_j|$.
Same and opposite sign vortices contribute differently to the correlations, a feature we will discuss in detail for few vortex states.
The normalized correlator reads
\begin{align}
    g^i(r)&=\frac{1}{\sum_{i=1,j=1}^Nq_iq_jG(0,r_{ij}/\xi)}\sum_{i=1,j=1}^Nq_iq_jG(r/\xi,r_{ij}/\xi),
    \label{gaccurate}
\end{align}
and has the properties:
\begin{enumerate}
    \item $g^i(0)=1$ by construction.
    \item For a single vortex we recover the correct expression \eref{Ge}.
    \item For $r\ll r_{ij}$, the correlator is dominated by the universal core contribution, and the point vortex far field is not relevant. 
    \item For $r\gg r_{ij}$, the point vortex far field dominates, and the total charge of the vortex distribution is relevant.
\end{enumerate}
We have developed a general treatment for distributions involving well-separated vortices, but the result does not affer any significant physical insight. To proceed further, we identify a useful approximation.
\subsection{Incompressible approximation}
Thus far our analysis is essentially exact, due to the accuracy of asymptotics used. Our aim is to retain the main compressible effects for $r\ll \xi$, but neglect the small compressible corrections around vortex cores at large distances. This will greatly simplify the analytical results. 

For the point vortex regime of well-separated vortices, since $\Lambda=0.8249\dots$, we always have $\Lambda r_{ij}\gg \xi$, and hence when $a \ll b$, we have $(\alpha,\beta)=(b,a)$, and it will be an excellent approximation in general that 
\begin{align}
    G(a,b)&\simeq \tilde G(a,b)\equiv I_0\left(\frac{\Gamma \alpha}{2}\right)K_0\left(\frac{\Gamma \alpha}{2}\right)-K_0(\Lambda \alpha ),
    \label{Uh}
    \end{align}
    provided $r_{ij}\gg \xi$. We refer to this as the incompressible approximation.
    The validity may be understood as a consequence of $\Gamma \ll \Lambda$, or $\xi\ll R$, as follows. On the face of it, one may guess that this is due to $I_0(x)\to 1$ for small arguments. However, it is valid more broadly as the second term is only a small correction to the first at large arguments due to its asymptotic suppression. The utility of this approach is shown by considering few-vortex special cases.

\subsection{Few vortex states}

\begin{figure}[!t]
    \centering
    \includegraphics[width=\columnwidth]{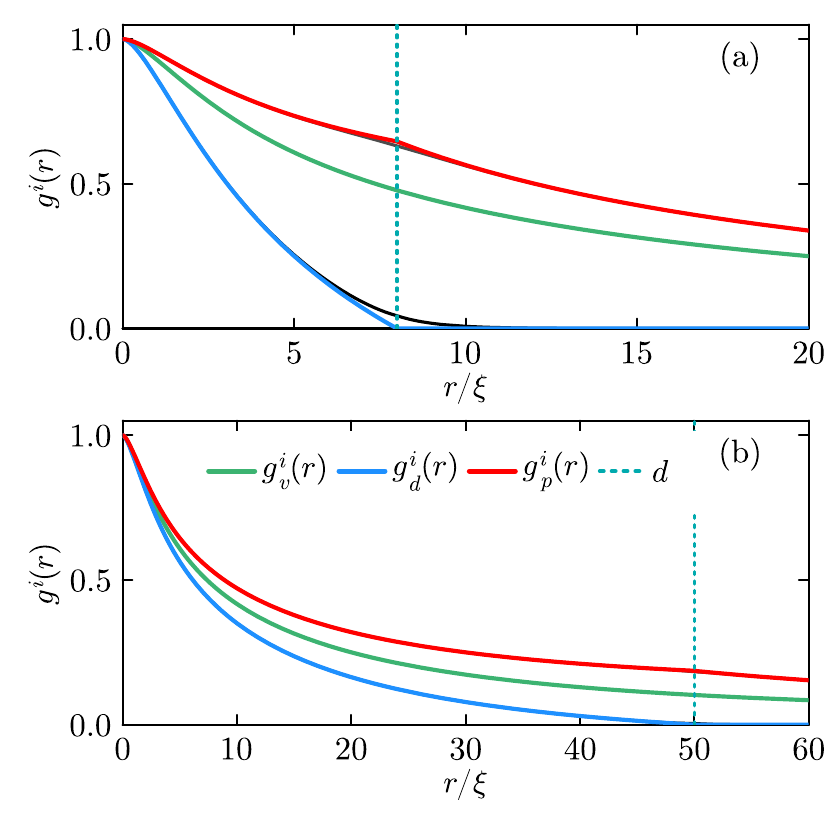}
    \caption{Two-point correlation functions of the incompressible velocity. The black lines show the (essentially exact) correlation function \eref{gaccurate}, and the colored lines show the approximate forms for a single vortex~\eref{gvone}, a vortex dipole~\eref{gvdip}, and a pair of same-sign vortices~\eref{gvpair}. (a) $d=8\xi$, (b) $d=50\xi$. For the infrared cutoff $\Gamma=2\pi/R$ we set $R=56\xi$ for both plots.}
    \label{fig:gih}
\end{figure}
We now consider simple few-vortex states using the incompressible approximation and find a more obvious statement about the role of velocity correlations.

\emph{Single vortex.---} To reiterate, a single vortex has two-point incompressible velocity correlator
        \begin{align}
            g_v^i(r) &\equiv \frac{K_0\left(\frac{\Gamma r}{2\xi}\right)I_0\left(\frac{\Gamma r}{2\xi}\right)-K_0\left(\frac{\Lambda r}{\xi}\right)}{\ln\left(\frac{2\Lambda}{\Gamma}\right)}.
            \label{gvone}
        \end{align}

\emph{Vortex Dipole.---} A pair of vortices of opposite sign separated by $d$ has correlator
\begin{align}
    g_d^i(r) &=\begin{cases}\dfrac{g_v^i(r)-g_v^i(d)}{1-g_v^i(d)},&r\leq d \\
        0,&r>d,
    \end{cases}
    \label{gvdip}
\end{align}
showing cancellation of velocity fields for small distances.

\emph{Vortex Pair.---} A pair of same sign vortices separared by $d$ has correlator
\begin{align}
    g_p^i(r) &=\begin{cases} \dfrac{g_v^i(r)+g_v^i(d)}{1+g_v^i(d)},&r\leq d\\
        \dfrac{2g_v^i(r)}{1+g_v^i(d)},&r>d,
        \end{cases}
        \label{gvpair}
\end{align}
where now the velocity field is enhanced at small distances, and decays like a single vortex at large distances.

Numerical example velocity correlations for the vortex, dipole, and pair are shown in \fref{fig:gih}. In (a) we show a small vortex separation, and in (b) a large separation. In contrast to the long range behavior of a single vortex, the dipole has velocity correlation that drops to zero near the separation scale $d$, corresponding to the significant cancellation of velocity at larger distances~\footnote{The local velocity correlation is finite outside the dipole separation scale, but vanishes under angular averaging.}. The vortex pair on the other hand has enhanced velocity correlation at short scales and at large scales decays like a charge 2 single vortex. Compressible effects are evident for (a) near the separation scale $d$, but are negligible in (b) where $\xi \ll d$.

Relative to the single vortex, the vortex pair correlator  at $r=d$ is enhanced by the factor $1< 2/(1+g_v^i(d))\leq 2$, due to the superposition of the two corotating vortex fields, as shown in (a) and (b). For larger $d$, the enhancement saturates near to a factor of 2, as shown in (b).
\section{Rarefaction pulse}
\label{sec:rarefaction}
We now turn to the case of a rarefaction pulse, which is a compressible solitary wave solution of the GPE. The rarefaction pulse appears naturally in the weak dissipative decay of a vortex dipole and it interesting to consider how the velocity correlations change under this transformation. To the best of our knowledge, the velocity power spectrum and two-point velocity correlation function for a rarefaction pulse are not known.

Starting from \eref{jrn}, \eref{jrp}, at leading order $\sqrt{n}v_x\simeq\sqrt{n_0}v_x$, i.e. the background density variations do not matter at leading order for a fast rarefaction pulse, and our aim is to understand the velocity field. Using the quantum phase \eref{jrp}, we immediately arrive at 
\begin{align}
    \sqrt{n_0}v_x&=\frac{\sqrt{n_0}\hbar}{m}(-2\sqrt{2}\epsilon)\partial_x\frac{\sqrt{2}\epsilon x/\xi}{3/2+2(\epsilon x/\xi)^2+2(\epsilon^2y/\xi)^2}.
\end{align}
The Fourier transform can then be written as
\begin{align}
    {\cal F}[\sqrt{n_0}v_x]&=\frac{\sqrt{n_0}\hbar}{m}(-2\sqrt{2}\epsilon)ik_x\notag\\
    &\times\frac{1}{2\pi}\int d^2\br\:e^{-i\bk\cdot\br}\frac{\sqrt{2}\epsilon x/\xi}{3/2+2(\epsilon x/\xi)^2+2\epsilon^2(\epsilon y/\xi)^2}.
\end{align}
We can try to proceed exactly, but the anisotropy means that the polar integral in $k$-space, $\int d\theta_k$ cannot be evaluated. We have identified a simple approximation that captures the velocity decorrelation length scale at short range: we simply ignore the slower decay in the $y$ direction of the denominator, setting $\epsilon^2\to 1$. Note that our calculation is equivalent to retaining only the leading term in an expansion in powers of $v^2/c^2<1$. Including higher orders can in principle be done (next order also captures the second much weaker peak in the spectrum, however the correlation function is no longer accessible analytically). Here we are only interested in the dominant length scale in the two-point correlator, captured by the leading term. 

We hence treat the denominator as cylindrically symmetric:
\begin{align}
    {\cal F}[\sqrt{n_0}v_x]&\simeq\frac{\sqrt{n_0}\hbar}{m}(-2\sqrt{2}\epsilon)ik_x\notag\\
    &\times\frac{1}{2\pi}\int d^2\br\:e^{-i\bk\cdot\br}\frac{\sqrt{2}\epsilon x/\xi}{3/2+2(\epsilon x/\xi)^2+2(\epsilon y/\xi)^2},
\end{align} 
but retains the dominant anisotropy due to the phase jump in the $x$ direction. This approximation captures well the main length scale of the velocity correlation function we are interested in. 

Changing variables $\sqrt{2}\epsilon x/\xi=\sqrt{3/2}u$, $\sqrt{2}\epsilon y/\xi=\sqrt{3/2}v$, we have $dxdy = (3/2)\xi^2/(2\epsilon^2) dudv$, and the integral becomes
\begin{align}
    {\cal F}[\sqrt{n_0}v_x]&\simeq\frac{\sqrt{n_0}\hbar}{m}(-\sqrt{3}\xi^2/\epsilon)ik_x\notag\\
    &\times\frac{1}{2\pi}\int dudv\:e^{-i(k_x u+k_yv)\xi\sqrt{3}/(2\epsilon)}\frac{u}{1+u^2+v^2}.
\end{align}
By first doing the polar integral using \eref{Kpol}, the integral can be evaluated to give 
\begin{align}
    {\cal F}[\sqrt{n_0}v_x]&\simeq i\frac{\sqrt{n_0}\hbar}{m}\frac{\sqrt{3}\xi^2}{\epsilon}\frac{k_x^2}{k}K_1(k\lambda/2),\\
    {\cal F}[\sqrt{n_0}v_y]&\simeq i\frac{\sqrt{n_0}\hbar}{m}\frac{\sqrt{3}\xi^2}{\epsilon}\frac{k_xk_y}{k}K_1(k\lambda/2).
\end{align}
where $\lambda\equiv\sqrt{3}\xi/\epsilon$ turns out to be the scale of the angle-integrated velocity correlation function. In this approximation, the (compressible) energy spectrum \eref{vspecdef} is 
\begin{align}
    E^c(k)&=\varepsilon_0\xi\frac{3}{2\epsilon^2}(k\xi)^3K_1(k\lambda/2)^2.
    \label{Ecr}
\end{align}
The two-point correlator is 
\begin{align}
    G(r)&=\int_0^\infty dk\:E(k)J_0(kr)\notag\\
    &=\varepsilon_0\xi^4\frac{3}{2\epsilon^2}\int_0^\infty dk\:k^3J_0(kr)K_1(k\lambda/2)^2\notag\\
&=\varepsilon_0\xi^2\int_0^\infty dz\:z^3J_0(zr/\lambda)K_1(z/2)^2,
\end{align}
where we changed to $z=k\lambda$.
Making use of integral \eref{J0K1}, we find the normalized correlator $g^c_\lambda(r)\equiv G(r)/G(0)\equiv f(r/\lambda)$, where
\begin{align}
    f(x)&=\frac{3}{8(x^2+1)^2}\left(1-\frac{1}{2x^2}+\frac{2(x^2+1/4)}{x^3\sqrt{x^2+1}}\operatorname{csch}^{-1}\left(\frac{1}{x}\right)\right).\label{twopt}
\end{align}
This function, and its dependence on $r/\lambda$, captures the key length scale of rarefaction pulse velocity correlations. An example is shown in the following section which also examines the effect of a background density gradient. For the localized structures of dipole and rarefaction pulse the background variation effect is minimal when the background is locally slowly varying.

It is worth noting that although the rarefaction pulse wavefunction has a lower velocity limit of $v\simeq 0.8c$ for physical validity (positive density), there is no such limit on the two point correlation function~\eref{twopt}. This property was recently used to characterize correlations near the point of rarefaction pulse formation~\cite{baker-rasooli_observation_2025}.
\section{Trapped system}\label{sec:trap}
While analytical treatments of the homogeneous system are useful, most systems of interest are inhomogeneous. In this section we compare our analytical results to spectra and correlations for excitations in the center of a trapped system, where the background density is inhomogeneous. We consider a harmonic trap containing a BEC in the Thomas-Fermi regime, and find numerically that the analytical results for the incompressible velocity power spectrum and two-point correlation function are remarkably accurate, with some additional finite size effects.

An inhomogenous background density is more commonly encountered in experiments, both in BECs and photon quantum fluids~\cite{fontaine_observation_2018,glorieux_hot_2023}. It is worthwhile to consider the role of background curvature of the quantum fluid density, and the new scale introduced by the system size. We consider a harmonic trap with frequency $\omega_\perp$, and work in units of the oscillator length $a_\perp=\sqrt{\hbar/m\omega_\perp}$. The system consists of a BEC with chemical potential $\mu=40\hbar\omega_\perp$, with healing length $\xi=0.158a_\perp$, and Thomas-Fermi radius $R=\sqrt{2\mu/m\omega_\perp}=56.6a_\perp$. We use a numerical grid of $(n_x,n_y)=(512,512)$ points, over a square domain of side length $5R$ to represent the trapped system as a Thomas-Fermi ground state. In trap units, the dimensionless interaction strength $\varg = 0.1$ corresponds to a system of $N_{a}=5\times 10^4$ atoms. 

\begin{figure}[!t]
    \centering
    \includegraphics[width=\columnwidth]{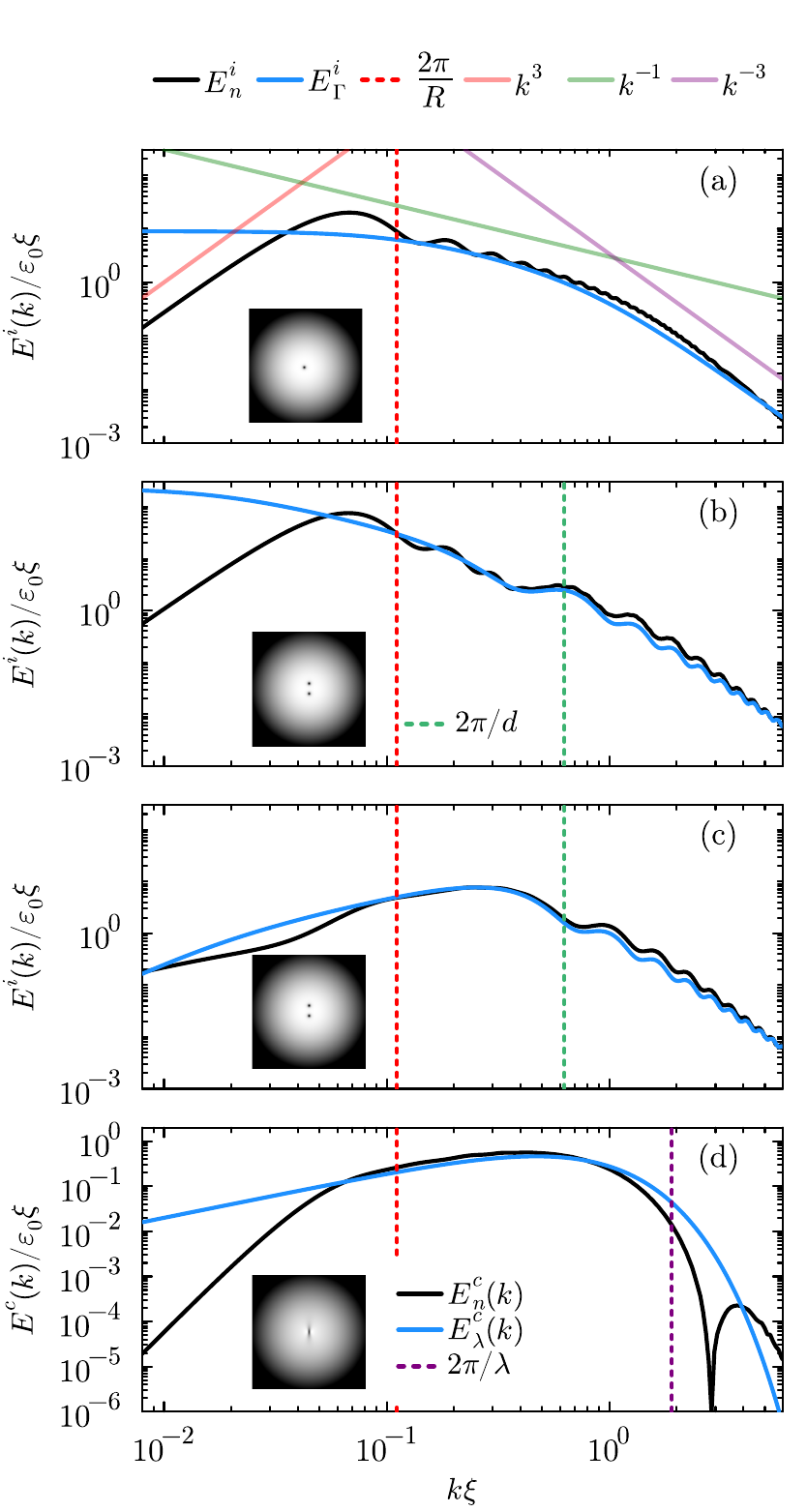}
    \caption{Energy spectra in a trapped BEC, comparing the numerical spectrum, $E_n^{(i,c)}(k)$, for a Thomas-Fermi state imprinted with (a) a single vortex (density inset), with power laws for comparison; (b) a pair of vortices of the same sign with separation $d=10\xi$; (c) a vortex dipole with the same separation; (d) Rarefaction pulse with $v=0.85c$. The numerically calculated spectra, $E_n^{i,c}$ are compared with \eref{Eivort} for (a)-(c), and \eref{Ecr} for (d), where $\lambda=\sqrt{3}\xi/\epsilon$. The analytical expressions are evaluated using an infrared cutoff $\Gamma=2\pi/R$, and the Thomas-Fermi radius is $R=56\xi$ for all subplots. The rarefaction pulse has velocity $v=0.85c$, giving $\epsilon=0.526$.}
    \label{fig:combs}
\end{figure}

Velocity power spectra and correlations are calculated numerically using the high resolution Fourier method described in~\cite{bradley_spectral_2022}, and summarized in Appendix \ref{app:c}. In essence, the problem of binning on coarse momentum grids is avoided by doing the angular integral in momentum space analytically. Each point in the numerical spectrum then requires a forward and inverse Fourier transform to complete, gaining flexibility and accuracy (especially for low $k$) in the spectrum. High resolution allows accurate numerical evaluation of \eref{Gdef}. The incompressible velocity power spectra are shown in \fref{fig:combs} for a single vortex, a pair of vortices, a vortex dipole, and a rarefaction pulse with $v=0.85c$. 

\begin{figure}[!t]
    \centering
    \includegraphics[width=\columnwidth]{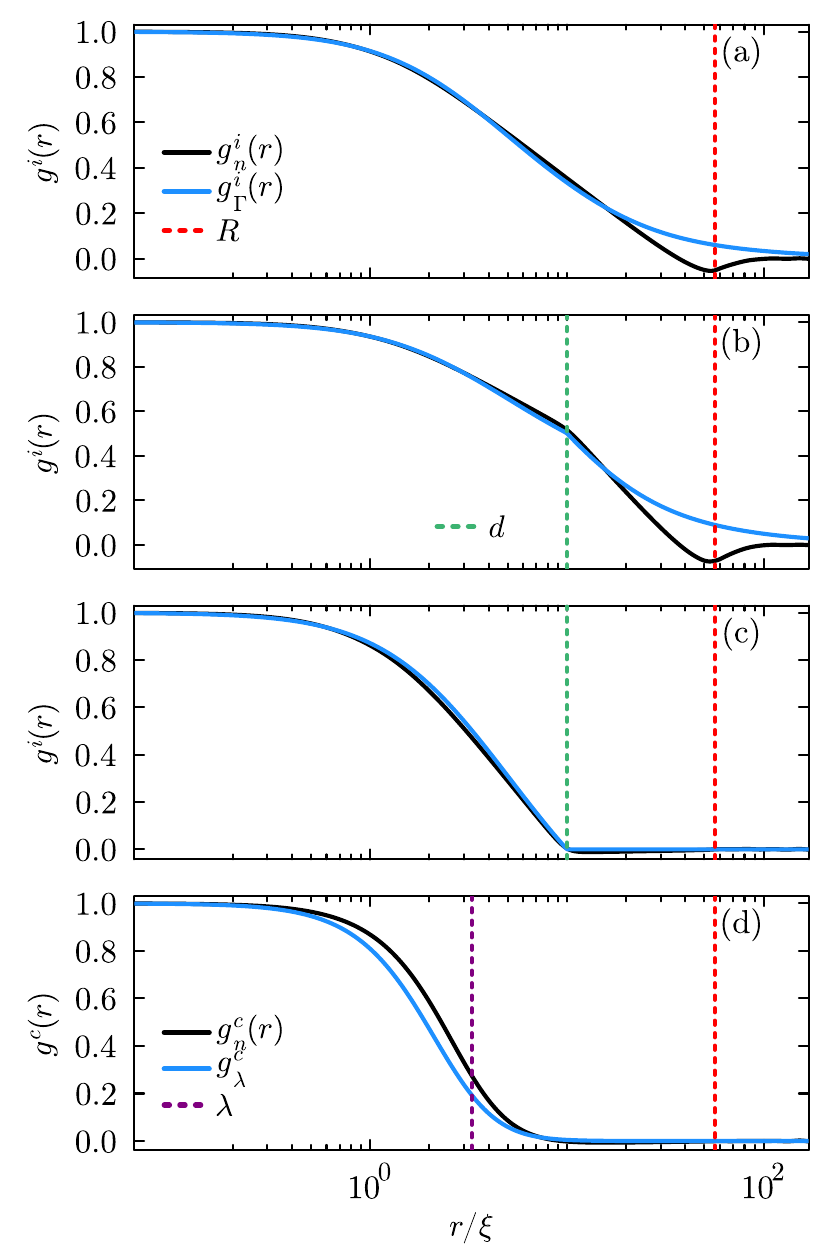}
    \caption{Angle averaged velocity two point correlation function, $g^{(c,i)}(r)$ in a trapped BEC. (a) a single central vortex. (b) a pair of vortices with separation $d=10\xi$. (c) a vortex dipole with the same separation; (a)-(d) show the incompressible velocity correlation function. (d) a rarefaction pulse with $v=0.85c$, showing the compressible velocity correlation function, with $\lambda=\sqrt{3}\xi/\epsilon$. The black lines show the numerically calculated correlation function for inhomogeneous system, and the blue lines show the analytical expressions for a homogeneous system with an infrared cutoff $\Gamma=2\pi/R$, with $R=56\xi$ for all subplots. The rarefaction pulse has velocity $v=0.85c$, giving $\epsilon=0.526$.}
    \label{fig:combg}
\end{figure}

The velocity correlation functions are shown in \fref{fig:combg}. For vortices, the characteristics are akin to \fref{fig:gih}, with finite size effects now evident at small $k$, corresponding to different long range decay of the velocity correlation at distances of order $R$; these differences are only strongly evident for the long-ranged configurations in \fref{fig:combg} (a), (b). For the vortex dipole and rarefaction pulse, \fref{fig:combg} (c), (d), we see the characteristic short range coherence lengths $d$ and $\lambda$ for compressible and incompressible velocity respectively. 

The trapped system results have recently been tested against a high resolution experiment in a quantum fluid of light, finding good quantitative agreement~\cite{baker-rasooli_observation_2025}. Such an analysis is only possible in a quantum fluid of light where the Helmholtz decomposition can be carried out on high resolution density and velocity fields~\cite{nore_kolmogorov_1997,bradley_spectral_2022}. 

\section{Conclusions}\label{sec:conclusions}
We developed an analytical framework to describe energy spectra and velocity correlations of quantum vortices and rarefaction pulses in planar quantum fluids. The interplay between kinetic and nonlinear energies determines the vortex core shape, but the exact form remains analytically unknown. To enable tractable calculations an analytical core profile is commonly assumed. In this work, we introduced a new vortex core ansatz with improved analytical integrability, allowing a more systematic treatment of vortices. Using this exponential ansatz, we solved the general problem of calculating the power spectrum of a planar compressible quantum fluid containing vortices. By Fourier transforming the spectrum, we derived an analytical expression for the two-point velocity correlator of vortex distributions.

We analyzed simple vortex distributions by applying an incompressible approximation, which revealed distinct signatures of short- and long-range velocity coherences---associated with vortex dipoles and vortex pairs, respectively---and their characteristic length scales. Vortex pairs exhibited enhanced correlation due to their coherent long-range velocity fields, whereas vortex dipoles showed velocity correlations that decayed to zero over the vortex separation scale. We compared these vortex results with a similar analysis of the Jones-Roberts rarefaction pulse, which also features a characteristic short-range velocity coherence length. Our analytical predictions align well with numerical treatment of a harmonically trapped system, where we imposed excitations at the center of a smoothly inhomogeneous background.
 
The exponential vortex ansatz may have a number of wider applications in the study of 2D quantum vortex dynamics~\cite{mehdi_mutual_2023} and quantum turbulence~\cite{bradley_energy_2012}. A future area of interest is the role of dissipative effects in rarefaction pulse dynamics~\cite{krause_thermal_2024}, for which velocity correlations could prove a useful characterization. Vortex core changes in dipolar systems~\cite{bland_vortices_2023}, and compressible effects in planar quantum turbulence~\cite{bradley_energy_2012,reeves_inverse_2013-1,reeves_identifying_2015} may also be an interesting direction for investigation. Generalized point vortex models that include dissipative effects~\cite{mehdi_mutual_2023}, or finite vortex mass~\cite{richaud_vortices_2020} could benefit from a more tractable vortex core ansatz. It may also be possible to extend the exponential vortex ansatz to three-dimensional excitations such as vortex lines and rings~\cite{svidzinsky_dynamics_2000,fetter_vortices_2001}.

\acknowledgements
AB would like to acknowledge the hospitality of Quentin Glorieux's research group at LKB ENS Sorbonne Universit\'{e} where this work was initiated. The authors acknowledge the Dodd-Walls Centre for Photonic and Quantum Technologies for financial support.
\appendix
\section{Spectral analysis for compressible quantum fluids}\label{app:c}
We give a short summary of the main result of Ref.~\cite{bradley_spectral_2022} which is used to numerically evaluate the velocity power spectra and two-point correlation functions presented in this work. The method aims for fast accurate Helmholtz decomposition of the velocity field, which allows separation of compressible and incompressible parts. The key innovation is carrying out angular integral to create the spectrum analytically. The remaining numerical problem is evaluated using fast Fourier transforms (FFTs). 

Here we summarize the main results. Given two complex valued vector fields $\uu$, $\vv$, we define the spectral density of their inner product, $\langle\uu\|\vv\rangle(k)$,  as the function that integrates to the inner product $\ip{\uu}{\vv}$ over all $k$-space, i.e.
\begin{align}\label{specdens_main}
    \langle \uu|\vv\rangle&\equiv\int_0^\infty dk\;\langle\uu\|\vv\rangle(k).
\end{align}
After some algebra, in $d$ spatial dimensions this can be written as 
\begin{align}\label{specdens_main}
    \langle\uu\|\vv\rangle(k)&\equiv\int d^d\x\;\Lambda_d(k,|\x|)C[\uu,\vv](\x),
\end{align}
where the two-point correlation in position space is
\begin{align}\label{Cdef_main}
    C[\uu,\vv](\x)&\equiv\int d^d\mathbf{R}\;\ip {\uu}{\mathbf{R}-\x/2}\ip{\mathbf{R}+\x/2}{\vv}, 
\end{align}
and the kernel function is dimension dependent:
\begin{align}\label{Lamdef_main}
    \Lambda_d(k,r)&\equiv
    \begin{cases}
        \frac{1}{2\pi}\;kJ_0(kr),&\text{for } d=2,\\
        \frac{1}{2\pi^2}\;k^2\sinc{(kr)},&\text{for } d=3.
    \end{cases}
\end{align}
The cartesian two-point correlation is thus directly mapped to a spectral density function of $k$ by analytically integrating over angles in $k$-space; the two point correlation function \eref{Cdef_main} may be efficiently evaluated using FFTs. Similarly, there is an exact mapping from $k$-space back to an angle-average of the two-point correlation function \eeref{Cdef_main}, in the form
\begin{align}\label{guv_main}
    g_{uv}(r)\equiv\int_0^\infty dk\;\Lambda_d^{-1}(k,r)\frac{\langle\uu\|\vv\rangle(k)}{\langle\uu|\vv\rangle },
\end{align}
with inverse kernel
\begin{align}\label{Laminvdef_main}
    \Lambda_d^{-1}(k,r)&\equiv\begin{cases}
        J_0(kr)&\text{for } d=2,\\
        \sinc{(kr)}&\text{for } d=3.
    \end{cases}
\end{align}
We have absorbed the value of $C[\uu,\vv](\mathbf{0})=\langle\uu |\vv\rangle$ for convenience, so that $g_{uv}(0)\equiv 1$ by definition.  We can summarize this as a statement of an angle-averaged Wiener-Khinchin theorem: $(\uu,\vv)\longrightarrow \langle \uu||\vv\rangle(k)\longleftrightarrow g_{uv}(r)$,
where the first mapping is a Fourier transform followed by angular integration in $k$ space. The second is the Fourier relation between the $k$ and $r$ variables. Our analysis amounts to an explicitly angle-averaged formulation of the standard Wiener-Khinchin theorem linking power spectra with two-point correlation functions. This formulation for general vector fields allows many quantities of interest to be precisely calculated for compressible quantum fluids, by choosing $\uu$ and $\vv$ appropriately.

\section{Bessel integrals}
We collect several useful Bessel function integrals and integral representations here for completeness. In some cases these are already tabulated, while others are less well known. 

For working with vortices we use
\begin{align}
    \int_0^{\infty} e^{-\alpha x} J_\nu(\beta x) d x&=\frac{\beta^{-\nu}\left[\sqrt{\alpha^2+\beta^2}-\alpha\right]^\nu}{\sqrt{\alpha^2+\beta^2}},
    \label{appj1}
\end{align}
\begin{align}
    \int_0^{\infty} x^{\nu+1} J_\nu(a x) \frac{d x}{x^2+b^2}&=b^\nu K_\nu(a b),
    \label{appj2}
\end{align}
\begin{align}
    \int_0^{\infty} \frac{J_0(a z) J_0(b z)z}{z^2+c^2}d z & =
    I_0(\text{min}(a,b) c) K_0(\text{max}(a,b) c)  
    \label{JJrat},
\end{align}
\begin{align}
    \frac{1}{\sqrt{z^2+\Gamma^2}}&=\frac{1}{\pi}\int_{-\infty}^\infty dk\:K_0(\Gamma |k|)e^{ikz}
    \label{FTK0},
\end{align}
\begin{align}
    J_0(x)&=\frac{1}{\pi}\int_{-1}^1\frac{e^{ikx}}{\sqrt{1-k^2}}dk
    \label{FrepJ},
\end{align}
\begin{align}
    \int_0^1 du \frac{K_0(au)}{\sqrt{1-u^2}}&=\frac{\pi}{2}I_0(a/2)K_0(a/2).\label{Kint}
\end{align}
For the rarefaction pulse we use the following  integrals 
\begin{align}
    \frac{1}{2\pi}\int dudv\:e^{-i(k_x u+k_yv)a}\frac{u}{1+u^2+v^2}=-iK_1(ak)k_x/k,
    \label{Kpol}
\end{align}
\begin{align}
    \int_0^\infty dk&\:k^3J_0(a k )K_1(b k/2)^2=\frac{4}{a^3(a^2+b^2)^2}\notag\\
    &\times\left(a^3-\frac{ab^2}{2}+\frac{2b^2(a^2+b^2/4)}{\sqrt{a^2+b^2}}\operatorname{csch}^{-1}\left(\frac{b}{a}\right)\right).
    \label{J0K1}
\end{align}

\section{Correlation integral}\label{appU}
To calculate the two-point correlation for vortices we require the integral
\begin{align}
    U(a,b)&=\int_0^\infty dz\:F_\Gamma(z)J_0(za)J_0(zb)\notag\\
    &=\int_0^\infty dz\:\left(\frac{1}{\sqrt{z^2+\Gamma^2}}-\frac{z}{z^2+\Lambda^2}\right)J_0(za)J_0(zb).
\end{align}
Note that $U(a,0) = G_\Gamma^i(a)$, and $U(0,0) = G_\Gamma^i(0)=\ln\left(2\Lambda/\Gamma\right)$.
The second term can be integrated using \eref{JJrat}. This is always well defined since $K_0$ is always evaluated at the larger argument, avoiding the logarithmic divergence for small arguments. 

The first integral
\begin{align}
    I(a,b,\Gamma)&=\int_0^\infty dz\:\frac{J_0(az)J_0(bz)}{\sqrt{z^2+\Gamma^2}}
\end{align}
is not known in closed form. However, we can evaluate it to a very good approximation as follows. First, we use the Fourier transform \eref{FTK0} and extend the $z$ integral to the real line
\begin{align}
    I(a,b,\Gamma)&=\frac{1}{2\pi}\int_{-\infty}^\infty dk\:K_0(\Gamma |k|)\int_{-\infty}^\infty dz\:J_0(az)J_0(bz)e^{ikz}.
\end{align}
Consider the case $a\gg b$, and put $u=az$. Then 
\begin{align}
    I(a,b,\Gamma)&=\frac{1}{2\pi a}\int_{-\infty}^\infty dk\:K_0(\Gamma |k|)\int_{-\infty}^\infty du\:J_0(u)J_0(bu/a)e^{iku/a}\notag\\
    &\simeq\frac{1}{\sqrt{2\pi} a}\int_{-\infty}^\infty dk\:K_0(\Gamma |k|)\int_{-\infty}^\infty du\:J_0(u)\frac{e^{iku/a}}{\sqrt{2\pi}}.
\end{align} 
Also, we use the Fourier representation of the Bessel function \eref{FrepJ}, or the Fourier transform pair 
\begin{align}
    J_0(x)&\longleftrightarrow\sqrt{\frac{2}{\pi}}\frac{\Pi(k/2)}{\sqrt{1-k^2}},
\end{align}
where $\Pi(x)$ is the rectangular function of unit width and height. We then have
\begin{align}
    I(a,b,\Gamma)&\simeq\frac{1}{\pi a}\int_{-\infty}^\infty dk\:K_0(\Gamma |k|)\frac{\Pi(k/2a)}{\sqrt{1-(k/a)^2}}\notag\\
    &=\frac{2}{\pi }\int_0^{1} du\:\frac{K_0(\Gamma ak)}{\sqrt{1-k^2}}.
\end{align}
Using the integral \eref{Kint} 
we have 
\begin{align}
    I(a,b,\Gamma)&= I_0(\Gamma a/2)K_0(\Gamma a/2)\quad \text{for} \quad a \gg b.
\end{align}
The integrand is symmetric in $a$ and $b$, and we require both limits. Since the two asymptotic results are continuous at $a=b$ we use it for all $a,b$ as a matched asymptotic expansion:
\begin{align}
    I(a,b,\Gamma)&=\begin{cases}
        I_0(\Gamma a/2)K_0(\Gamma a/2)\quad \text{for} \quad a >b,\\
        I_0(\Gamma b/2)K_0(\Gamma b/2)\quad \text{for} \quad a <b.
    \end{cases} 
\end{align}
This turns out to be an excellent approximation for all $a,b$ of interest, and is used in the main text.
Putting it together, we have the transform 
\begin{align}
    U(a,b)&=\int_0^\infty dz\:F_\Gamma(z)J_0(za)J_0(zb)\notag\\
    &=I(a,b,\Gamma)-\int_0^\infty dz \frac{z}{z^2+\Lambda^2}J_0(za)J_0(zb)\notag\\
    &=\begin{cases}
        I_0(\Gamma a/2)K_0(\Gamma a/2)-K_0(\Lambda a )I_0(\Lambda b ) \quad \text{for} \quad a>b,\\
        I_0(\Gamma b/2)K_0(\Gamma b/2)-K_0(\Lambda b )I_0(\Lambda a ) \quad \text{for} \quad a<b.
        \end{cases}
\end{align}
  


\providecommand{\noopsort}[1]{}

\end{document}